\documentclass[sigconf]{acmart}
\setcopyright{none}  
\renewcommand\footnotetextcopyrightpermission[1]{}
\acmConference[SenSys '26]{Proceedings of the 22nd ACM Conference on Embedded Networked Sensor Systems}{May 2026}{Saint-Malo, France}
\acmYear{2026}
\copyrightyear{2026}
\acmDOI{10.1145/XXXXXXX.XXXXXXX} % TODO: set upon acceptance
\acmISBN{978-1-4503-XXXX-X/26/11}
\acmBooktitle{Proceedings of the 22nd ACM Conference on Embedded Networked Sensor Systems (SenSys '26), May 2026, Saint-Malo, France}

\usepackage{balance}
\usepackage{enumitem}

\title{Lessons Learned from Developing a Privacy-Preserving Multimodal Wearable for Local Voice-and-Vision Inference}

\author{Yonatan Tussa}
\affiliation{%
  \institution{University of Maryland, College Park}
  \country{USA}
}
\email{ytussa@umd.edu}

\author{Andy Heredia}
\affiliation{%
  \institution{University of Maryland Global Campus}
  \country{USA}
}
\email{aheredia7@umgc.edu}

\author{Nirupam Roy}
\affiliation{%
  \institution{University of Maryland, College Park}
  \country{USA}
}
\email{niruroy@umd.edu}

\begin{document}

\begin{abstract}
Many promising applications of multimodal wearables require continuous sensing and heavy computation, yet users reject such devices due to privacy concerns. This paper shares our experiences building an ear-mounted voice-and-vision wearable that performs local AI inference using a paired smartphone as a trusted personal edge. We describe the hardware-software co-design of this privacy-preserving system, including challenges in integrating a camera, microphone, and speaker within a 30-gram form factor, enabling wake word-triggered capture, and running quantized vision-language and large-language models entirely offline. Through iterative prototyping, we identify design hurdles in power budgeting, connectivity, latency, and social acceptability. Our initial evaluation shows that fully local multimodal inference is feasible on commodity mobile hardware with interactive latency. We conclude with design lessons for researchers developing embedded AI systems that balance privacy, responsiveness, and usability in everyday settings.
\end{abstract}

\maketitle

% Hero Image - removed for now
% \begin{figure}[t]
%   \centering
%   \includegraphics[width=\linewidth]{fig_hero}
%   \caption{Device overview: wearable earpiece, paired smartphone, and local AI inference.}
%   \label{fig:hero}
% \end{figure}
  
\section{Introduction}

Wearable devices promise contextual intelligence, but achieving this vision requires always-available sensing and computation. Continuous data capture and cloud inference have fueled privacy backlash, leading to the rejection of devices such as camera glasses and body-mounted loggers~\cite{Hoyle2014PrivacyBehaviorsLifeloggers}. Prior work on privacy-aware lifelogging and visual privacy protection~\cite{Korayem2014ScreenAvoider, shu2016cardeacontextawarevisualprivacy} highlights how user discomfort often stems from ambiguity about when recording occurs and how data is used. These lessons remain highly relevant as new wearable devices emerge, embedding microphones, cameras, and sensors near the head for continuous multimodal interaction~\cite{Hu2025EarableSurvey, Fan2023DesignEarableSensing, Elfouly2025WearableComputingComprehensiveSurvey}.

At the same time, advances in mobile accelerators and compact multimodal models make it increasingly feasible to perform meaningful inference locally, without uploading raw data to external servers. 
Our work explores this convergence: a lightweight, privacy-preserving form factor that supports large-model reasoning for assistive tasks such as situational awareness (``what am I looking at?''), conversational assistance, and potentially memory support. 
We examine the system-level challenges of enabling large-model reasoning under the strict constraints of a small, user-facing device—balancing power, latency, comfort, and trust while keeping computation local to the user's own devices. 
By demonstrating that local multimodal inference is practical within these limits, this work lays the groundwork for applications in \emph{memory augmentation, contextual recall, and personalized assistive intelligence}.

This paper contributes:
\begin{enumerate}
    \item \textbf{Experiences from building a privacy-preserving multimodal wearable}, including lessons in hardware integration, power management, network configuration, and sensor placement within a small, heat-constrained form factor.
    \item \textbf{Insights from implementing local large-model inference} on personal smartphones, detailing quantization, communication, and latency trade-offs for multimodal pipelines.
    \item \textbf{Early observations and evaluation plans}, exploring how privacy and form factor shape device perception.
\end{enumerate}

\section{Background and Related Work}

Wearable sensing systems have long aimed to provide continuous, context-sensitive intelligence, but early lifelogging work revealed persistent challenges around privacy, social acceptability, and ambiguity in recording practices. Continuous first-person capture raises concerns for both wearers and bystanders, with perceived sensitivity strongly shaped by location, activity, and who or what appears in view \cite{Hoyle2014PrivacyBehaviorsLifeloggers, Yoon2017EthicalUseLifelogging, Sellen2010BeyondTotalRecall}. Empirical reviews of deployments in health and everyday settings catalog practical precautions and residual risks associated with wearable cameras \cite{Meyer2022WearableCamerasEthics, Meyer2023WearableCamerasDailyLife}, while recent analyses show how egocentric video can leak personal traits well beyond user intent \cite{Li2025EgoPrivacy, article}. Complementary technical defenses include redacting sensitive on-screen content \cite{Korayem2014ScreenAvoider}, providing context-aware protections for bystanders \cite{Shu2016Cardea}, and mechanisms that allow people or environments to signal opt-out preferences \cite{Raval2016MarkIt, Denning2014SmartphoneCamera}. Our design responds to this body of work by defaulting to short, event-driven capture and confining all processing to user-owned devices.

A parallel thread studies memory augmentation and just-in-time information retrieval. Early systems such as Forget-Me-Not and other “memory prostheses” explored continuous capture and retrieval of everyday experiences \cite{Lamming1994ForgetMeNot, Starner1998AugmentedMemory, Sellen2010BeyondTotalRecall}. More recent systems move toward lightweight, context-aware assistance: Memoro uses large language models (LLMs) to infer users’ memory needs from conversational context and surface concise suggestions through an audio wearable \cite{Zulfikar_2024}, while Scribe demonstrates simultaneous voice and in-air handwriting interaction for voice assistants \cite{10.1145/3631411}. These efforts highlight both the promise of real-time cognitive support and the risk of disruption when wearable interfaces demand too much attention. Our work takes inspiration from this line of research but focuses on multimodal sensing (voice and vision) in an ear-mounted form factor, with a design goal of minimizing disruption and preserving social acceptability.

At the same time, \emph{earable} systems are rapidly maturing as a head-adjacent sensing and interaction platform. Surveys of ear-based technology document sensing modalities (audio, inertial, physiological), form-factor constraints, and emerging social norms around ear-mounted devices \cite{Hu2025EarableSurvey, Elfouly2025WearableComputingComprehensiveSurvey}. Industry perspectives emphasize that comfort, antenna placement, and the legibility of device behavior are central to adoption in everyday settings \cite{Fan2023DesignEarableSensing}. Acoustic and ultrasonic sensing work such as BackDoor and related systems show that microphones near the head can be leveraged for rich interaction and sensing without adding large new hardware \cite{Roy2017BackDoor}. Our prototype draws on these lessons, adopting a low-profile ear-mounted sensor that offloads heavy computation to a paired smartphone rather than embedding a full AI stack in the wearable itself.

Advances in \emph{on-device multimodal inference} further reshape the design space. Modern smartphones now include NPUs and GPUs capable of running compact language and vision-language models locally, including Qwen2-VL–style architectures optimized for edge deployment \cite{Qwen2VL2024} and FastVLM-style pipelines that target mobile and embedded platforms \cite{FastVLM2025}. Lightweight visual encoders such as FastViT reduce memory and compute requirements for mobile vision workloads \cite{Vasu2023FastViT}. Prior work on mobile and embedded ML has also explored software and model-level optimizations, from DeepX and DeepSense for resource-efficient inference on phones and sensor data \cite{Lane2015DeepX, Huynh2017DeepSense} to DeepIoT and integer-only quantization schemes tailored to low-power devices \cite{Yao2017DeepIoT, Jacob2018Quantization}. We build on these developments via MLX \cite{mlx2023} and Apple’s on-device ASR framework \cite{AppleSFSpeechRecognizer} to support fully offline speech, language, and vision-language inference, with all sensor data remaining on the user’s smartphone.

For wake word detection, keyword spotting (KWS) on microcontrollers is backed by TinyML literature. The Speech Commands dataset provides a widely used benchmark for limited-vocabulary speech recognition \cite{Warden2018SpeechCommands}, while convolutional architectures such as small-footprint CNNs, TC-ResNet variants, and microcontroller-focused models trade accuracy against memory and latency \cite{Sainath2015KWS, Choi2019TCResNet, Zhang2017HelloEdgeKWS}. Our wake word model follows this line of work but emphasizes pragmatic choices to close the training–deployment gap, including synthetic-voice augmentation and hardware re-recording on the actual device microphone.

Finally, efficient \emph{on-device routing and intent classification} is well supported by classic text-categorization techniques. TF–IDF features combined with linear models remain strong baselines for short utterances \cite{Sebastiani2002TextCategorization, Pedregosa2011ScikitLearn}. In our system, such lightweight classifiers route queries to device control, visual, or conversational pipelines, reserving heavier LLM/VLM inference for the final stage.

Prior work highlights three constraints: (i) privacy and social acceptability concerns around continuous capture, (ii) the tight thermal and power budgets of ear-mounted hardware, and (iii) the emerging feasibility of fully local multimodal inference on smartphones. These threads motivate the central questions of our work:
\begin{itemize}[leftmargin=*, itemsep=0pt, topsep=2pt]
    \item Can an ear-mounted device with minimal compute and battery deliver responsive multimodal assistance when paired with a smartphone?
    \item How should sensing and connectivity be structured to balance bandwidth, latency, and energy cost?
    \item What properties are required for such a device to feel privacy-preserving and socially acceptable in everyday use?
\end{itemize}

Our answers are grounded in the practical experience of building and iterating on a real, functioning prototype.

\begin{figure}[t]
  \centering
  \includegraphics[width=0.48\linewidth]{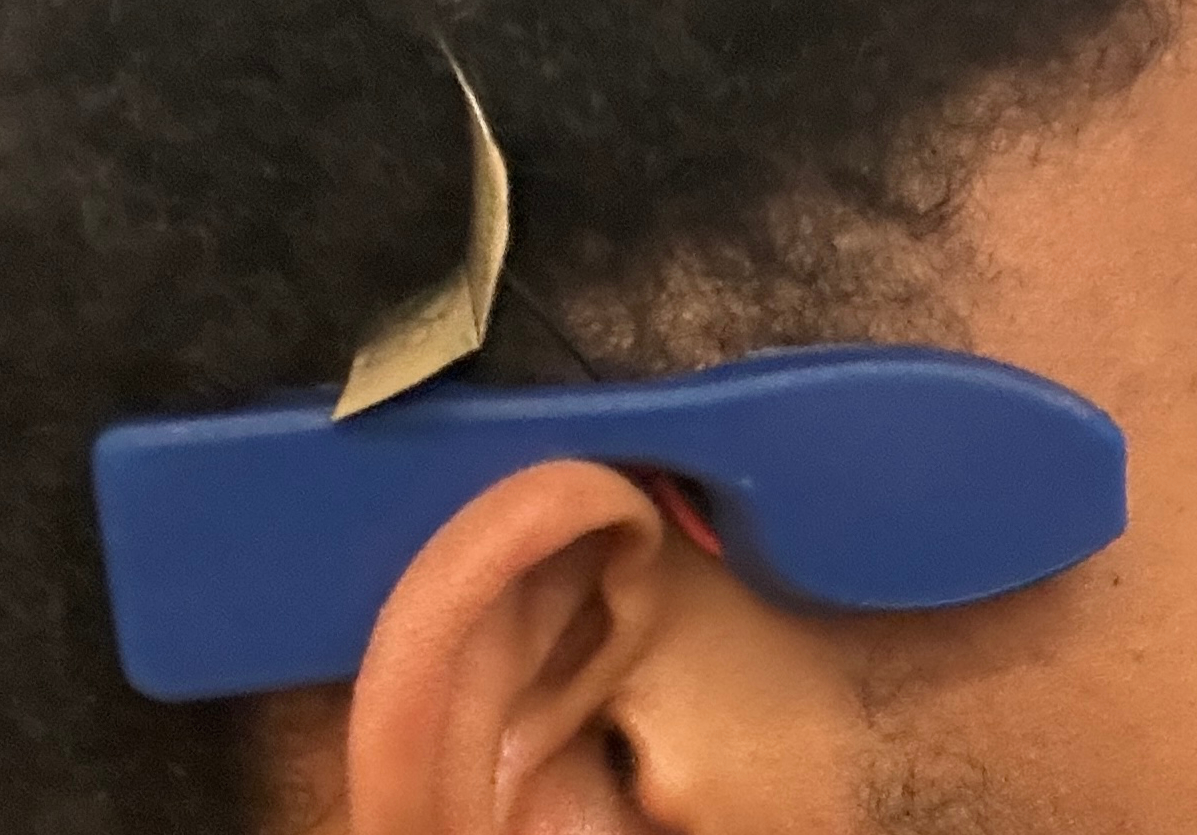}
  \hfill
  \includegraphics[width=0.48\linewidth]{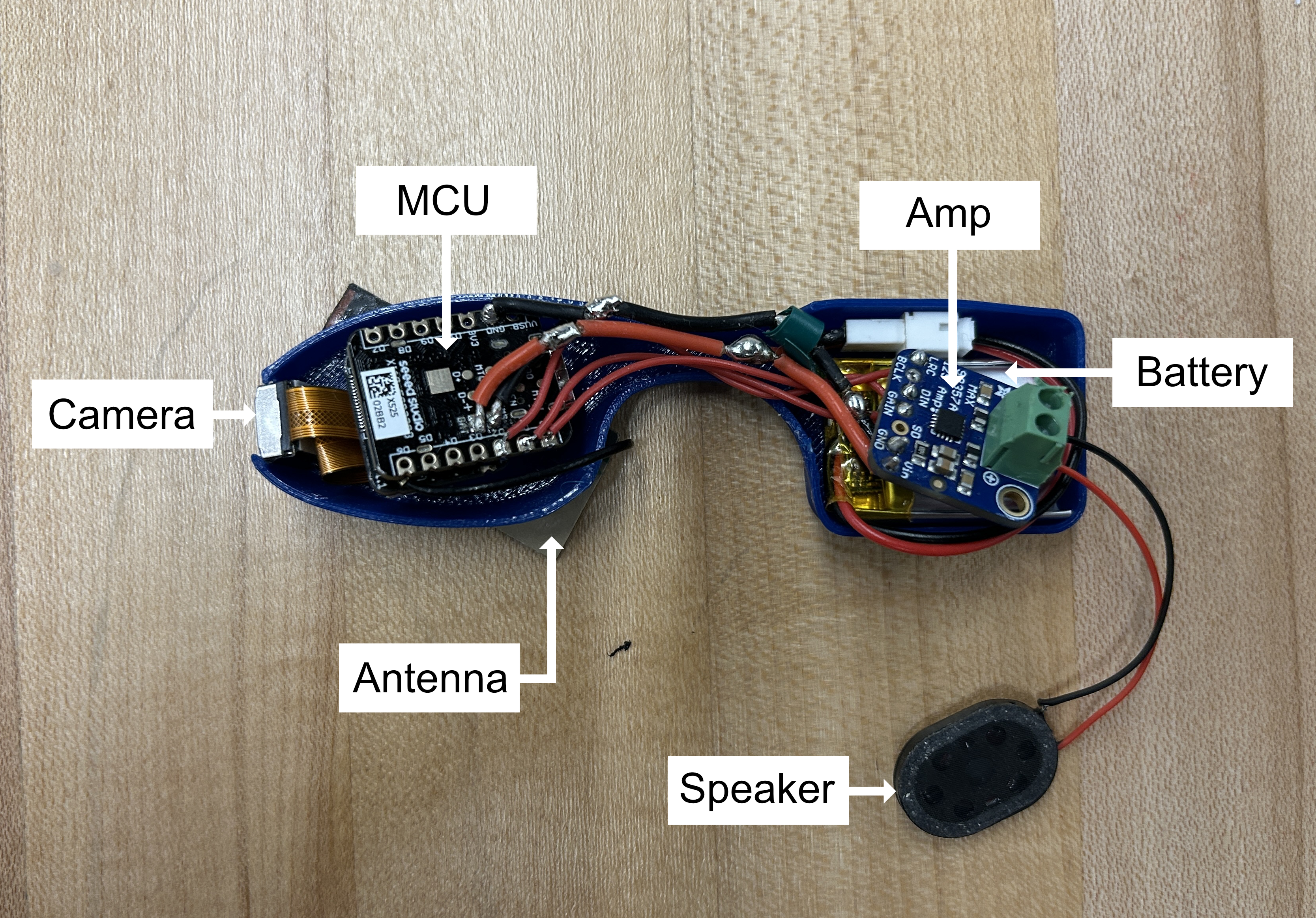}

  \caption{(a) Ear-mounted prototype. The device integrates a camera, microphone, speaker, MCU, and battery in a pen-shaped enclosure that is worn over the ear. (b) Close-up view of the prototype with internal components exposed, showing the camera module, battery, and MCU within the housing.}
  \label{fig:prototype}
\end{figure}

\section{Experiences Building the Device}

\begin{figure*}[t]
  \centering
  \includegraphics[width=\textwidth]{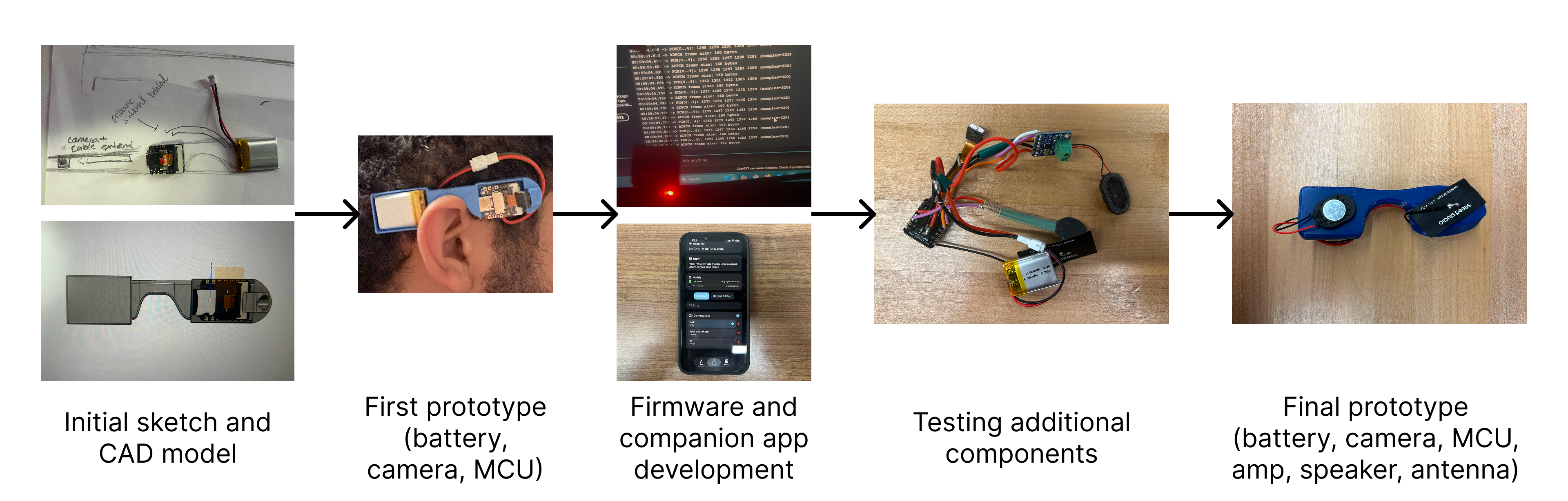}
  \caption{Iterative hardware–software development process of the device. Each step addressed issues in connectivity, ergonomics, and firmware stability.}
  \label{fig:process}
\end{figure*}

\subsection{Hardware Integration}

The device integrates a small camera (OV5460 AF),  microphone (MEMS from MCU) , speaker (1W, $8 \Omega$) paired with amp (Class D, 3W, $8-4 \Omega$), and a Wi-Fi/BLE-capable microcontroller (XIAO ESP32S3 Sense) inside a 3D-printed enclosure that hooks over the ear (Figure~\ref{fig:prototype}). The camera near the temple approximates the user's field of view. A 200~mAh Li-ion battery along the behind-ear segment powers the device. The total weight is approximately 30~g. BLE is used solely for device discovery and secure provisioning (e.g., setting Wi-Fi credentials). After setup, all runtime communication (audio, discrete photo snapshots, and control messages) runs over Wi-Fi/mobile hotspot, which offers sufficient bandwidth and more predictable latency for interactive inference. We fabricated three functional prototypes with this architecture, experimenting with different components and enclosure materials: PLA, resin, TPU (Figure~\ref{fig:prototype}).

\textit{Bring-up via a minimal webserver}. A turning point in the hardware-software co-design was treating the MCU as a standalone webserver during bring-up. Before attempting wake word detection or multimodal streaming, we deployed a minimal TCP WebSocket server derived from the \textit{CameraWebServer} example in the XIAO ESP32S3 Sense documentation. We exposed endpoints for toggling an on-board LED, capturing a single camera frame, and sending/receiving short audio buffers. Interacting with these endpoints via simple web requests (e.g., \texttt{curl}) let us rapidly validate power behavior, connectivity, and sensor functionality. This also forced us to resolve early system details before layering on the full inference pipeline, including IP advertising, Wi-Fi reconnection after sleep, and control message formats.

\textit{Mechanical design and ergonomics}. Packing the camera, battery, MCU, and speaker into an ear-mounted form required several design iterations. We experimented with camera positioning—ultimately angling it just behind the temple to balance field-of-view and visual conspicuity—and tested multiple hook shapes for fit and stability. Using 3D-printed prototypes on a reference head model, we identified a gentle concave hook contour that “catches” behind the ear without requiring a custom fit. We refined the curvature and thickness until the device remained secure even during vigorous motion or downward-facing poses. These adjustments produced a pen-like silhouette that follows the head’s contour—more akin to a familiar ear accessory than a protruding camera—which improved both comfort and social acceptability.

\textit{Connectivity robustness}.
Users naturally move between access points and through partial occlusions, so rather than tuning low-level TCP behavior we structured communication around short sessions and compact payloads. The device streams 4-bit IMA-ADPCM audio to the companion app, achieving a $\sim$4$\times$ size reduction over 16-bit PCM. The app decodes frames, reconstructs a continuous waveform for ASR, and symmetrically re-encodes synthesized responses before sending them back to the device.

Bringing up this pipeline required several iterations. Early unidirectional streams yielded clicks and pops caused by encoding artifacts and buffer under-runs; switching to a ring buffer with $\sim$300 ms of play-out delay eliminated these issues while keeping latency acceptable. Adding the reverse audio path exposed new failures--repeated samples, degraded playback, and occasional stalls--that initially appeared to be decoder bugs. Instead, they stemmed from Wi-Fi/BLE coexistence: both radios share the 2.4 GHz path, and bidirectional audio stressed the automatic coexistence policy. Explicitly prioritizing BLE only during provisioning and Wi-Fi during normal operation resolved these instabilities without hardware changes.

% \begin{figure}[t]
%   \centering
%   \includegraphics[width=0.7\linewidth]{fig2_prototype}
%   \caption{ TODO: (Placeholder for now, add image of internals with labels) Close-up view of the prototype with internal components exposed, showing the camera module, battery, and MCU integration within the ear-mounted enclosure.}
%   \label{fig:prototype_internals}
% \end{figure}

\subsection{Power Budgeting and Battery Life}

Our ear-mounted prototype uses a 200\,mAh Li-ion cell to power a XIAO ESP32S3 Sense, Wi-Fi/BLE radios, and a 3W Class D amplifier driving a 1W
speaker. Board measurements and vendor data indicate that light-sleep draws only a
few milliamps, modem-sleep with the radio ready consumes on the order of
25\,mA, and deep-sleep falls to tens of microamps. Recording audio increases
board current to roughly 50-60\,mA, while enabling Wi-Fi or BLE in active mode
pushes consumption to about 80-100\,mA. The on-board camera is particularly
expensive: capturing VGA-resolution frames can add hundreds of milliamps in short
bursts. The Class D audio amplifier contributes only a few milliamps of
quiescent current when idle, but during 1\,W playback its input power corresponds
to roughly 300\,mA at 4\,V.

With a 200\,mAh cell, ``all-day'' operation imposes a hard constraint: averaged
over 24 hours the device must draw no more than about 8\,mA. The rough budget in
Table~\ref{tab:power-budget} shows that this is incompatible with keeping the
board in a constant ``always-listening + BLE advertising'' state.

\begin{table}[t]
  \centering
  \small
  \caption{Approximate current draw for operating states (200\,mAh cell).}
  \label{tab:power-budget}
  % slightly tighter horizontal padding
  \setlength{\tabcolsep}{2pt}
  \begin{tabular}{p{0.18\columnwidth}p{0.46\columnwidth}p{0.16\columnwidth}p{0.16\columnwidth}}
    \hline
    \textbf{Scenario} & \textbf{Components on} & \textbf{Approx.\ current} & \textbf{Approx.\ run-time} \\
    \hline
    Baseline (listening for wake word) &
    MCU in modem-sleep, mic + MFCC pipeline armed for wake word,
    BLE advertising at low duty cycle, amplifier in standby &
    $\approx$80-100\,mA &
    $\approx$2\,h \\
    \hline
    Active query burst &
    Wi-Fi TX/RX, camera capture, mic streaming, amplifier playing back
    speech, control logic active &
    $\approx$400-450\,mA &
    $\approx$25-30\,min (if continuous) \\
    \hline
  \end{tabular}
\end{table}

In practice, active queries last only a few seconds, so these high-power episodes
contribute relatively little energy compared to the cost of keeping the wake word
path armed. If the device were to run in ``always listening'' mode continuously,
the 200\,mAh cell would be exhausted after roughly two hours. To approach
day-scale usage, the system must therefore spend most of its time in deep or
light sleep and only enable wake word listening in short, user-initiated
sessions (for example, a 30-minute walking or commuting interval). This
trade-off, using a small, lightweight battery versus continuous availability ultimately
led us to design the current prototype for short interaction sessions rather than
true 24-hour standby.

\subsection{Local Inference Pipeline}

\begin{figure}[t]
  \centering
  \includegraphics[width=\linewidth]{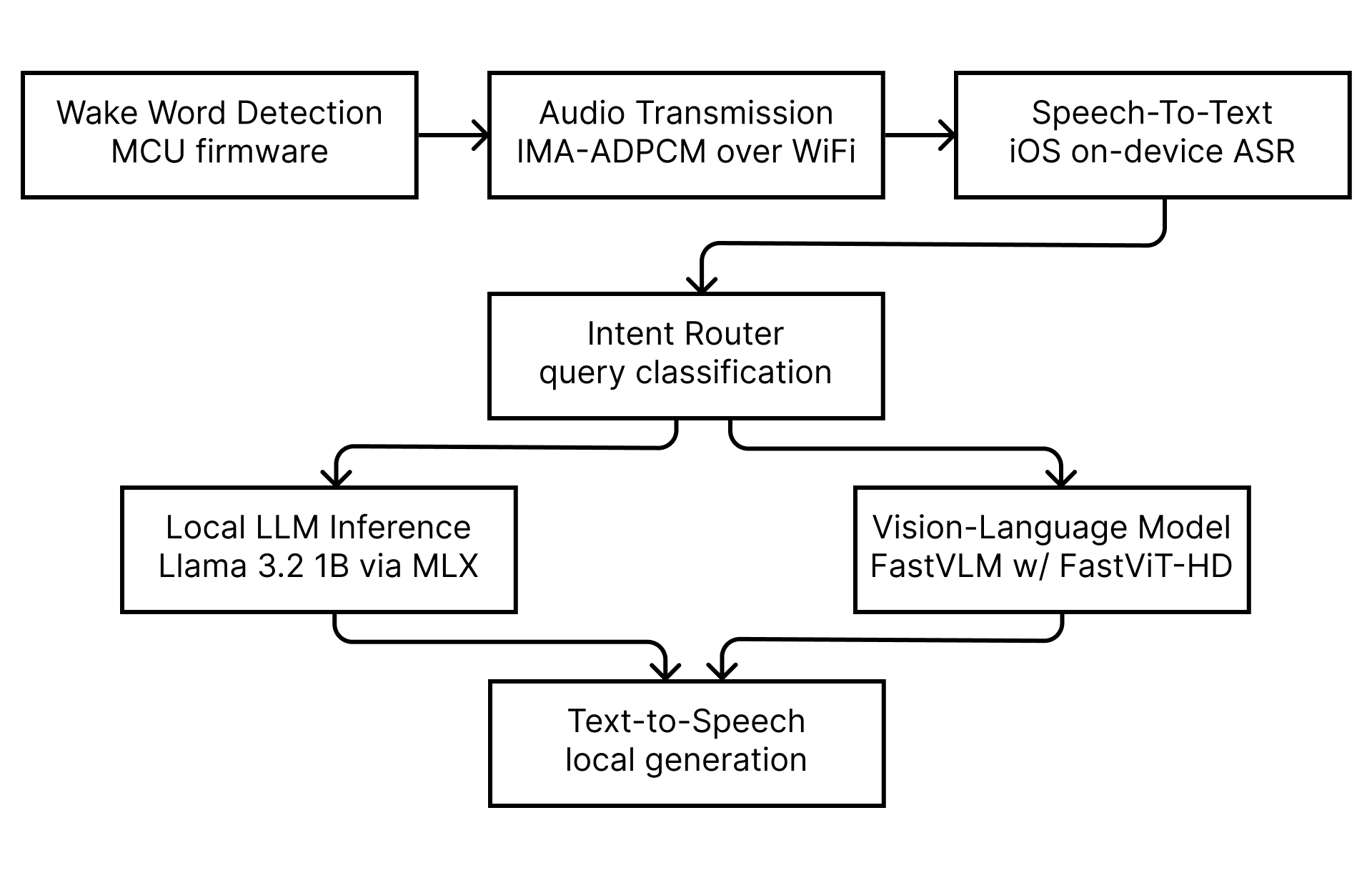}
  \caption{Local inference pipeline showing the flow from wake word detection to local multimodal inference and response generation. All processing executes locally on the smartphone.}
  \label{fig:pipeline}
\end{figure}

Triggered by a wake word, the device records the user's query audio and, for vision-grounded queries, triggers a single photo capture. Both are sent over Wi-Fi to the paired smartphone, which executes the entire inference pipeline locally using Core ML and MLX. Figure~\ref{fig:pipeline} shows the overall inference flow.

\paragraph{Audio capture and preprocessing.}
The firmware streams IMA-ADPCM compressed audio chunks to the phone. The app decodes them to 16~kHz mono PCM, applies clamping and DC offset removal, and routes the cleaned signal to ASR.

\paragraph{Speech-to-text.}
We use Apple's on-device \texttt{SFSpeechRecognizer}, configured for streaming transcription without network access. Partial hypotheses are available as the user speaks; silence detection marks the end of an utterance and re-arms wake word detection on the device.

\paragraph{Intent routing.}
The text transcript is passed to an intent router that identifies command-like queries (e.g., ``take a picture'') versus open-ended or visual questions (e.g., ``what's on this table?''), and dispatches either to simple handlers or to multimodal inference.

\paragraph{Local language-model inference.}
Conversational queries run on a locally hosted LLaMA~3.2 1B model, quantized to 4-bit and served via MLX with Metal GPU acceleration. Inference is gated on foreground app state to avoid iOS background GPU issues; GPU caches are cleared between calls, using 128-384~MB depending on device.

\paragraph{Vision-language inference.}
For visual queries, the captured frame is processed by FastVLM, a Core~ML implementation of a lightweight Qwen2VL-derived model with a FastViT-HD encoder. The model outputs scene descriptions, OCR-like text extraction, and simple question-answering, returned as structured JSON for downstream logic.

\paragraph{Text-to-speech and streaming back.}
Responses are synthesized locally using \texttt{AVSpeechSynthesizer}. A parallel path captures the waveform, resamples to 16~kHz mono, applies IMA-ADPCM compression, and streams 320-sample chunks back to the earpiece speaker. All audio and visual data remain confined to the user's phone.

\paragraph{Privacy and performance.}
The pipeline operates without cloud calls or third-party servers. Adaptive GPU cache sizing, lazy model loading, and compressed audio streaming keep latency low while respecting memory constraints. Internal tests on an iPhone~14 show end-to-end latency of roughly 2-3 seconds for typical queries.

\subsection{Intelligence Architecture}

Beyond the sensor and network layers, the companion app implements
a hierarchical intelligence architecture that governs query routing,
model selection, and privacy enforcement.
This layer bridges lightweight on-device classifiers with heavier
local inference models, ensuring responsiveness without relying on
cloud services.

\paragraph{Intent classification and routing.}
Incoming user queries—typically transcribed speech from the wake word pipeline—
first pass through a lightweight \emph{intent classifier} implemented
using a TF-IDF + Logistic Regression model. This model runs entirely
on-device and routes text into one of four categories:
\texttt{device\_control}, \texttt{visual\_query}, \texttt{general\_question},
or \texttt{conversational}. The classifier executes within tens of milliseconds
and determines which downstream processing path to trigger:
direct device commands (e.g., “take a photo”), visual queries invoking
the FastVLM model, or conversational responses handled by the
LLaMA~3.2~1B language model.  
This simple text-based classifier serves as the entry point of the
system’s intelligence hierarchy and minimizes latency by filtering
non-LLM interactions early.

\paragraph{Routing and model orchestration.}
The \emph{intent router} manages pipeline execution after classification.
It orchestrates three main pathways:
(1) device control, which issues commands directly to the earpiece firmware;
(2) visual queries, which perform image capture followed by FastVLM
analysis; and (3) conversational queries, which invoke the language
model with optional memory retrieval. The router enforces strict
foreground execution and model loading to uphold privacy guarantees.

\paragraph{Local inference and privacy enforcement.}
The heavier models in the pipeline--LLaMA~3.2~1B (quantized to 4-bit)
for text and FastVLM (Qwen2-VL) for vision-language understanding--are loaded dynamically only while the app is in the foreground.
Inference is performed fully locally using Apple’s MLX and Core~ML
frameworks, ensuring no data leaves the user’s device. Intermediate
results (transcripts, embeddings, and scene descriptors) are stored in
encrypted Core~Data containers, and the app maintains a zero-API-call
policy for all intelligence functions.

\paragraph{Future extensions.}
Two additional classifiers---a \emph{memory need classifier} and a
\emph{moment importance classifier}---are under development.
The former determines when to retrieve prior interactions or captured
moments for contextual augmentation, while the latter evaluates the
significance of captured scenes to decide whether they should be
archived. These components form the foundation for more proactive
and memory-aware interactions in future iterations of the system.

\begin{figure}[t]
  \centering
  
  \includegraphics[width=\linewidth]{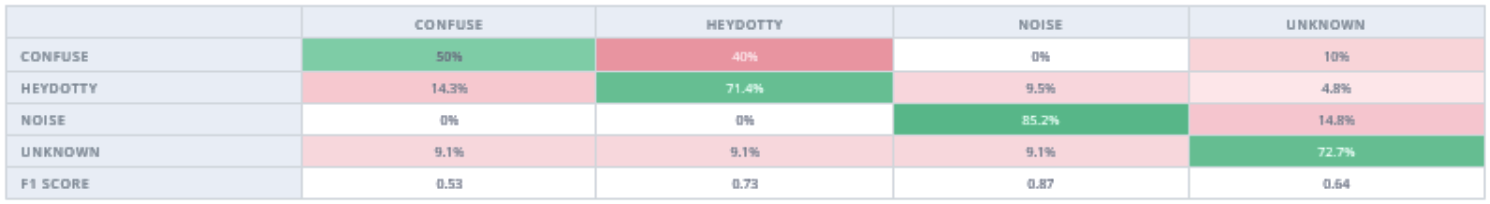}
  \vspace{0.5em}
  \includegraphics[width=\linewidth]{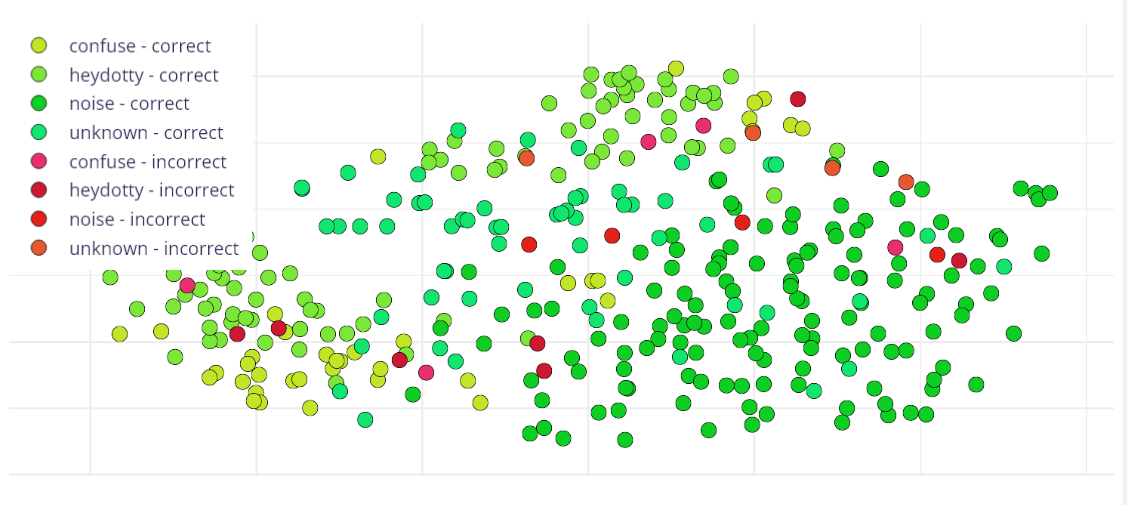}
  \caption{Wake word dataset visualization. 
  Top: class distribution across positive, confuser, and background samples. 
  Bottom: representative spectrograms and signal features recorded through the device microphone.}
  \label{fig:wake_dataset}
\end{figure}

\subsection{Wake word Dataset Design}
% \begin{itemize}
Reliable wake word detection on a small MCU required a dataset that captured how people might naturally say the chosen phrase across voices, speaking styles, and environments, while also covering common confusions. The phrase of choice is arbitrary, and for this prototype we went with "hey dotty". To bootstrap this dataset without large numbers of speakers, we used a commercial neural text-to-speech (TTS) system to synthesize diverse, high-quality training utterances. 

\textit{Synthetic wake word positives}. We curated roughly 50 synthetic voices (roughly balanced across perceived male and female timbres) and manually filtered out any that sounded obviously robotic. For each retained voice we generated multiple renditions of the wake word. These renditions varied in pitch, speaking rate (0.6-1.0 s duration), and prosody (different intonations/deliveries)  across synthesized utterances.

Before fixing the synthesis settings, we had someone repeatedly speak the wake word aloud in different imagined contexts (calling out loudly vs. quietly addressing the device) to calibrate what “natural” usage would sound like; this informed which TTS voices and prosody patterns were kept. 

\textit{Re-recording through the device}. Rather than train directly on clean TTS waveforms, we played each generated utterance through a mono speaker and re-recorded it with the actual device. The earpiece was worn and positioned with the microphone facing away from the speaker, on the “far side” of a head mock-up so the captured signal reflected the real acoustic path and microphone characteristics. We repeated this in several ambient environments (e.g., quiet room, office-style background chatter, white-noise ventilation, cafe-like ambiance, and other playlists of ambient noise) as well as in near-silence. For each voice and environment we aimed to collect multiple clean renditions of the wake word, yielding on the order of 20 usable samples per voice and a total of several hundred short clips (around 7 minutes of labeled audio). 

\textit{Negative and confuser classes}. To reduce false positives from phonetically similar phrases, we created a dedicated confuser class containing utterances such as “hey dobby,” and other near-rhymes. Additional random negatives consisted of short arbitrary phrases and clipped fragments (e.g., only “hey d-” or “-dotty”) that resulted when the wake word straddled a recording boundary. Finally, we added a noise class of environment-only audio with no foreground speech. Collecting these classes via the same re-recording setup ensured that microphone response and room acoustics were consistent across positives and negatives.

\textit{Iterating on segmentation}. Our initial collection script stored one-second WAV files to an SD card and required the speaker to say the word inside each interval. In practice this led to many clips where the word was partially cut at the start or end. Midway through development we switched to continuous recording and performed segmentation offline using a waveform viewer: we visually identified energy spikes corresponding to the wake word, previewed them, and exported tightly trimmed segments. This change significantly improved dataset quality at the cost of some manual curation. 

\textit{Data organization}. All final clips were stored as WAV files organized into label-based folders (heydotty, confuse, noise, unknown/random) to match the expected input format of our training pipeline. This structure also simplified later experimentation with class rebalancing (e.g., oversampling confusers to further reduce false triggers). Although we ultimately used only a subset of the available synthetic voices (10 male, 7 female) due to time constraints, the process is easily extensible to a larger pool in future iterations.

Figure~\ref{fig:wake_dataset} summarizes the wake word dataset
composition and illustrates the balance across positive, confuser,
and background classes, as well as example spectrograms recorded
through the device microphone. This visualization guided our
threshold selection and class-balancing experiments during model
training.

This synthetic-heavy but hardware-recorded dataset allowed us to quickly explore model architectures and thresholds on the MCU. In future work we plan to augment it with real user recordings, but this approach was sufficient to obtain a robust baseline wake word model for early prototypes.

\subsection{Training and Profiling}

To design a model that fits the constraints of our ear-mounted device, training and profiling was done using a commercial TinyML platform, Edge Impulse, targeting reference microcontrollers comparable to the XIAO ESP32-S3 Sense used in our prototype. The wake word detector operates on 1-second windows of 16 kHz mono audio with a 500 ms stride, producing overlapping predictions suitable for continuous listening.

Each audio window was converted into a 637-dimensional feature vector using Mel-Frequency Cepstral Coefficients (MFCCs) with 13 coefficients, a 25 ms frame length, 20 ms stride, 32 mel filters, and a pre-emphasis coefficient of 0.98. These features were tuned using Edge Impulse’s built-in “autotune” recommendations and qualitative inspection of feature maps to ensure clear separation between the wake word and background noise.

The final architecture was a compact 1-D convolutional neural network comprising two convolution-pooling blocks (8 and 16 filters, kernel size 3) with dropout (0.25) between layers, followed by flattening and a dense softmax layer with four outputs (heydotty, confuse, noise, unknown). We trained for 100 epochs using a learning rate of 0.005, batch size 32, and an 80/20 train-validation split. Edge Impulse’s INT8 quantization converted weights and activations to 8-bit precision for TensorFlow Lite Micro deployment.

The quantized model achieved 73.9 \% validation accuracy (AUC = 0.92), with the strongest separation between the wake word and background noise. Remaining confusions occurred mainly between “hey dotty” and near-rhymes such as “hey dobby,” which we treated conservatively as non-triggers. Despite the synthetic nature of the dataset, this performance was sufficient for real-time triggering on-device.

We used the platform’s built-in profiling tools to evaluate latency and resource use. On an 80 MHz reference microcontroller, total per-window latency was $\approx$ 370 ms (361 ms MFCC + 9 ms classification) with 15.4 KB peak RAM usage and 52 KB flash footprint. These measurements confirmed that the model fits comfortably within the ESP32-S3’s budget even when clocked below its 240 MHz maximum. Operating at 80 MHz in “always-listening” mode reduces static power and thermal load, and the $\approx$ 370 ms latency remains acceptable given that the spoken wake word itself lasts roughly one second.

To minimize false activations, we selected an operating point favoring a low false-activation rate (FAR $\approx$ 0.43 threshold, short averaging window $\approx$ tens of milliseconds, suppression window $\approx$ 1 s). This configuration prioritizes user trust and energy efficiency—missed detections can simply be repeated, whereas false triggers cause unnecessary inference and power drain. Overall, the resulting wake word detector provides a good balance of accuracy, latency, and energy cost suitable for continuous low-power monitoring on a tiny MCU.

\section{DISCUSSION}
Our experience highlights that building a privacy-preserving multimodal wearable is as much a systems integration problem as a modeling one. The feasibility of the device rested on aligning sensing, connectivity, and interaction design under strict constraints.

\textbf{Responsiveness on a constrained form factor (RQ1).} Achieving interactive latency within a small ear-mounted device required restricting the wearable to wake word detection, capture, and streaming while offloading all inference to a paired smartphone.  This allowed us to keep the device lightweight and thermally safe. Despite limited battery capacity and compute, we achieved query latencies of 2–3~s, showing that fully local multimodal interaction is feasible without cloud support.

\textbf{Balancing energy, latency, and bandwidth (RQ2).} Our design adopts a hybrid sensing pipeline: BLE for provisioning, Wi-Fi for runtime streaming, and short-lived event-driven sessions to reduce idle drain. Audio is compressed via 4-bit ADPCM, and single-frame snapshots replace continuous video. These decisions reduced power and network load while maintaining interactive performance.

\textbf{Designing for privacy and acceptability (RQ3).} Three properties shaped trust: (i) interaction is explicitly triggered by the user via wake word, (ii) all sensor data is processed locally, with no cloud calls, and (iii) the form factor is subtle and familiar. This design made the device feel unobtrusive in public. Still, some ambiguity remained; future iterations may include subtle LED or haptic cues to clarify capture status for both wearer and bystanders.

\textbf{Model pragmatism. }While the broader AI literature emphasizes scaling, we found that latency, memory, and thermal headroom mattered more. Compact, quantized models—especially when paired with dynamic model loading and cache tuning—enabled real-time interaction on commodity mobile hardware.

\textbf{System co-design.} Many integration challenges (e.g., Wi-Fi/BLE coexistence, buffer stability, antenna layout) only emerged late in iteration. Prototyping tools like the early webserver example allowed us to validate hardware, streaming paths, and thermal behavior before inference was integrated.

\subsection{Limitations}

Despite these advances, several limitations remain. The evaluation involved a small number of short-duration deployments and relied primarily on internal testing. Longer, in-the-wild studies are necessary to assess reliability, comfort, and social perception over time. Our analysis also focused on a single smartphone class (iPhone 14); behavior on lower-end hardware or across operating systems may differ. Finally, while the current prototype supports wake word activation and multimodal inference, it does not yet integrate higher-level capabilities such as contextual memory or continuous dialogue—areas we plan to address next.

\subsection{Future Work}

We are in the process of obtaining IRB approval for a pilot study to evaluate everyday usability. We plan to collect both quantitative and qualitative data, including end-to-end
latency, wake word accuracy per-participant, Wi-Fi reconnection behavior, battery
drop over sessions, and participant feedback on comfort, perceived
privacy, and bystander reactions. Future work will expand both the scope and depth of the system. On the hardware and networking side, we plan extended deployments, adaptive power management, and modular enclosures to support additional sensors and continuous use. Alternatives to wake word activation will also be explored further, including integration of a tactile sensor. On the intelligence layer, we plan to complete development of additional on-device classifiers for memory need and moment importance, enabling selective recall and local summarization. Finally, we plan to investigate human-facing aspects including social acceptability, accessibility, and personalization, supported by open-source release of our datasets and design toolkit for community replication.

\section{Conclusion}

We presented a privacy-preserving multimodal wearable that performs fully local voice-and-vision inference by leveraging a paired smartphone as a trusted edge. Through iterative hardware and software co-design, we demonstrated that it is possible to combine on-device wake word detection, local language and vision models, and low-latency communication within a 30-gram ear-mounted form factor. Our findings reveal three core design principles for future systems: event-driven sensing as the default, smartphone-as-edge as the architectural baseline, and model pragmatism over maximalism. Together, these principles outline a practical pathway for building embedded AI devices that respect both user trust and energy constraints. Beyond this prototype, we envision a broader class of wearable devices that use local AI not only to answer questions but to support memory, accessibility, and situational awareness in ways that remain private and socially acceptable.

\balance
\bibliographystyle{acm}
\bibliography{references}

\end{document}